\documentclass[lettersize,journal]{IEEEtran}
\usepackage{amsmath,amsfonts}
\usepackage{algorithmic}
\usepackage{algorithm}
\usepackage{array}
\usepackage[caption=false,font=footnotesize,labelfont=rm,textfont=rm]{subfig}
\usepackage{textcomp}
\usepackage{stfloats}
\usepackage{url}
\usepackage{verbatim}
\usepackage{graphicx}
\usepackage{cite}
\usepackage{amssymb}
\usepackage{hyperref}
\usepackage{dsfont}
\usepackage{diagbox}
\usepackage{array}
\usepackage{microtype}
\usepackage{color}
\usepackage{balance}
\usepackage{cuted}
\usepackage{setspace}
\usepackage{hyperref}
\hypersetup{colorlinks=true, linkcolor=blue, anchorcolor=blue, citecolor=blue}

\begin{document}
\title{ Ambiguity-Free Broadband DOA Estimation Relying on Parameterized Time-Frequency Transform}

\author{Wei Wang, Shefeng Yan,~\IEEEmembership{Senior Member,~IEEE,} Linlin Mao,~\IEEEmembership{Member,~IEEE,}\\ Zeping Sui,~\IEEEmembership{Member,~IEEE,} and Jirui Yang
\thanks{This work was supported in part by the National Natural Science Foundation of China under Grant 62192711 and Grant 62371447. \emph{(Corresponding author: Shefeng Yan.)}}

\thanks{W. Wang, S. Yan and J. Yang are with the Institute of Acoustics, Chinese Academy of Sciences, Beijing 100190, China, and also with the University of Chinese Academy of Sciences, Beijing 100049, China (e-mail: wangwei2022@mail.ioa.ac.cn; sfyan@ieee.org; yangjirui@mail.ioa.ac.cn).}

\thanks{L. Mao is with the Institute of Acoustics, Chinese Academy of Sciences, Beijing 100190, China (e-mail: maoll@mail.ioa.ac.cn).}

\thanks{Z. Sui is with the School of Computer Science and Electronics Engi
neering, University of Essex, CO4 3SQ Colchester, U.K. (e-mail: zepingsui@outlook.com).}\vspace{-2.5em}}



\maketitle

\begin{abstract}
\textls[-10]{An ambiguity-free direction-of-arrival (DOA) estimation scheme is proposed for sparse uniform linear arrays under low signal-to-noise ratios (SNRs) and non-stationary broadband signals. First, for achieving better DOA estimation performance at low SNRs while using non-stationary signals compared to the conventional frequency-difference (FD) paradigms, we propose parameterized time-frequency transform-based FD processing. Then, the unambiguous compressive FD beamforming is conceived to compensate the resolution loss induced by difference operation. Finally, we further derive a coarse-to-fine histogram statistics scheme to alleviate the perturbation in compressive FD beamforming with good DOA estimation accuracy. Simulation results demonstrate the superior performance of our proposed algorithm regarding robustness, resolution, and DOA estimation accuracy.}
\end{abstract}


\section{Introduction}
\textls[-10]{Direction-of-arrival (DOA) estimation has been extensively explored in various fields {such as wireless communication, radar and sonar systems} \cite{ref1,ref2,ref3,ref4}, and gained significant attention in the context of sparse arrays \cite{ref6}. However, since the {sparse array} invokes spatial undersampling, its performance suffers a lot due to DOA ambiguity \cite{ref7,ref8}. In contrast to the classic solutions that design array structures to mitigate the ambiguity of DOA \cite{ref9,ref10}, {frequency-difference (FD) manipulation can shift the beamformer to a lower out-of-band frequency with ambiguity-free DOA estimation \cite{ref26,ref27}}. Then, to further enhance the angular resolution in {FD-based conventional beamforming (FD-CBF) \cite{ref11}, FD-based multiple signal classification (FD-MUSIC) \cite{ref12}, compressive FD beamforming (CFD) \cite{ref13} and FD-based atomic norm minimization (FD-ANM) \cite{refr1}} have been proposed, {where the CFD exhibits super resolution but it relies on high signal-to-noise ratios (SNRs) \cite{ref13,ref26}}. 

Since most signals in active detection schemes are non-stationary broadband signals \cite{ref14,ref15}, the time-frequency analysis methods have been introduced to improve the performance of array signal processing \cite{ref16}, such as short-time Fourier transform (STFT) \cite{ref17} and Wigner-Ville distribution \cite{ref20}. Nonetheless, since the Heisenberg uncertainty principle limits the STFT-based concentration, its DOA estimation performance degrades significantly at low SNRs. Moreover, the cross terms in {Wigner-Ville distribution} lead to low DOA estimation accuracy when the array data extracts {time-frequency} samplings from cross terms. {Therefore, the MUSIC has been combined with parameterized time-frequency transform (PTFT), which provides higher {time-frequency} concentration than STFT via an internally defined kernel function without cross terms \cite{ref21}}. To the best of our knowledge, {time-frequency} analysis methods have not been exploited for ambiguity-free DOA estimation yet. {Thus, we conceive the PTFT-based unambiguous DOA estimation for non-stationary broadband signals.} {In addition, incoherently averaging DOA spectra is invoked to obtain the final DOA estimates, whereas the incoherent {averaging} cannot suppress the cross-term effect induced by FD processing in disturbed environments \cite{ref22}}.

Aiming at the above concerns, we conceive {a histogram statistics-aided compressive frequency-difference beamforming (HS-CFD) relying on PTFT} to strike ambiguity-free
DOA estimation for sparse uniform linear arrays. Our contributions are described explicitly below.
\begin{itemize}
	\item[$\bullet$] We conceive {the PTFT-based FD} to {achieve better signal energy aggregation compared to the conventional FD}, and its output SNR is compared with other counterparts in theory and simulation. {Then, the PTFT-based CFD is derived.}
	\item[$\bullet$] A coarse-to-fine histogram statistics technique is conceived {to replace conventional incoherent averaging}, in which grid searching is executed coarsely and finely to alleviate perturbation and preserve DOA estimation accuracy.
	\item[$\bullet$] Simulation results validate the above analysis results and demonstrate that the proposed algorithm can strike satisfactory robustness, high resolution, and superior accuracy compared to other counterparts.
\end{itemize}

{\textit{Notations:}} $(\cdot)^{\top}$, $(\cdot)^\ast$, $|\cdot|$, $\odot$ and $\lfloor\cdot\rfloor$ are transpose, complex conjugation, modulus, Hadamard product and rounding down, respectively. {$\int(\cdot)$ denotes an indefinite integral.} {$\angle(\cdot)$ computes the phase angle, e.g., $\angle e^{j\theta} = \theta$.} ${j} = \sqrt{-1}$ is an imaginary unit. $\lVert\cdot\rVert_p$ denotes $\ell_p$-norm term.

\section{{Problem Formulation}}
\label{problem}
\subsection{{Data Model}}

Assuming that an active detection system transmits a {non-stationary broadband linear frequency modulation} signal $s(t)$ with start frequency $f_{\rm L}$, end frequency $f_{\rm H}$, bandwidth $B$, {pulse duration $T$ and overall signal length $T_\mathsf{all}$}.
Consider $K$ far-field reflected signals impinge upon a { $M$-sensor array with spacing} $d>c/(2f_{\rm L})$.
{The frequency-domain data model} is written as
\begin{equation}
	\pmb{y}({f})=\sum_{k=1}^{K}\pmb{a}(f,\theta_{k})b_kS(f)e^{-{j}2\pi f\tau^k}+\pmb{n}(f),
	\label{3}
\end{equation}
where 
\begin{equation}
	\begin{aligned}		
        &\pmb{y}({f})=\left [Y_{1}(f),Y_{2}(f),\cdots,Y_{M}(f) \right ]^{\top},\\
        &\pmb{a}(f,\theta_{k})=\left [1, e^{-{ j}2\pi f\tau_{2}(\theta_{k})},\cdots,
		e^{-{j}2\pi f\tau_{M}(\theta_{k})} \right ]^{\top},\\
  &\pmb{n}(f)=\left[N_{1}(f),N_{2}(f),\cdots,N_{M}(f)\right]^{\top}. 
	\end{aligned}
 \nonumber
\end{equation}
{$Y_m(f),S(f),N_m(f)$} corresponds to the frequency spectra of the received signal $y_m(t)$, the original signal $s(t)$, and the noise signal $n_m(t)$ on the $m$th sensor. 
$b_k$ is the complex amplitude, $\theta_{k}\in[-90^{\circ},90^{\circ}]$, ${\tau^k}$ denotes the time delay between the $k$th target and the first sensor. and $\tau_{m}(\theta_{k})=(m-1)d\sin(\theta_{k})/c$ with $c$ being the propagation speed. $\{n_{m}(t)\}^M_{m=1}$ are {independent and identically distributed additive Gaussian white noise}. 
\vspace{-0.6cm}
\subsection{{CFD in \cite{ref13}}}
{Processed by FD, the received array data can be recast as }
\begin{equation}
    \begin{aligned}
        {\pmb{z}(f_1,f_2)=\pmb{y}({f_2})
			\odot\pmb{y}^\ast ({f_1}),}
    \end{aligned}
    \label{req1}
\end{equation}
{with $f_2=f_1+\Delta f$. Then, the CFD beamforming is derived by}
\begin{equation}
	\begin{aligned}
		{\pmb{x} = \underset{\pmb{x}\in\mathbb{C}^{N_{\rm grid}}}{\arg\min}\, \lVert\pmb{z}(f_1,f_2)-\pmb{A}({\Delta f})\pmb{x}\rVert^2_2+\mu\lVert\pmb{x}\rVert_1,}
		\label{req2}
	\end{aligned}
\end{equation}
{where $\pmb{x}$ denotes the solution that draws a DOA spectrum, $\pmb{A}({\Delta f})=\left[ \pmb{a}({\Delta f},\theta_1), \pmb{a}({\Delta f},\theta_2),\cdots,\pmb{a}({\Delta f},\theta_{N_{\rm grid}})\right]$ is a sensing matrix with $N_{\rm grid}$ discretized grids and $\mu>0$ is a regularization parameter.
$\pmb{a}(\Delta f,\theta)$ in $\pmb{A}({\Delta f})$ is given by $\pmb{a}({\Delta f},\theta)=\pmb{a}({f_{2}},\theta)\odot\pmb{a}^\ast({f_{1}},\theta)$.}
The resolution of beamforming improves with $\Delta f$ increasing \cite{ref11}, but the maximum $\Delta f$ is limited to $ c/(2d)$ for ambiguity-free DOA estimation \cite{ref13}. Besides, $(K^2-K)$ cross terms inevitably produce artifact DOAs \cite{ref12}. Also, it is intractable to accumulate the energy of non-stationary signals via fast Fourier transform (FFT). Hence, we propose the {PTFT-based HS-CFD} algorithm {involving} the {PTFT-based CFD} technique and the coarse-to-fine {histogram statistics} scheme. 
\vspace{-0.3cm}

\section{Proposed Method}
\label{proposed}
\subsection{{PTFT-based CFD}}
Upon exploiting PTFT to $s(t)$, we can obtain {\cite{rref1}}
\begin{equation}
	g_{s}(t,f_{w})=\int_{-\infty}^{\infty}S(f)\Gamma^{\rm R}(f)\Gamma^{\rm S}(f)H_{\sigma}^{\ast}(f-f_{w})e^{{j}2\pi ft}{\rm d}f,
	\label{7}
\end{equation}
where $\Gamma^{\rm R}(f)=e^{-{j}2\pi \int \kappa(f){\rm d}f}$, $\Gamma^{\rm S}(f)=e^{{j}2\pi f \kappa(f_{w})}$ and $\kappa(f)$ is the kernel function to be defined later. $H_{\sigma}(f-f_{w})$ denotes the $w$th analysis window with center frequency $f_{\rm L}\leq f_{w}\leq f_{\rm H}$, and a window width of $\sigma$. For the sake of demonstration, a rectangular window function is exploited, yielding
\begin{equation}
	{H_\sigma(f-f_w)=}
	\left\{
	\begin{aligned}
		1,&\ f_w-\frac{\sigma}{2}\leq f\leq f_w+\frac{\sigma}{2},\\
		0,&\ {\rm otherwise}.\\
	\end{aligned}
	\right.
	\label{8}
\end{equation}
Then, the stationary phase method is utilized to obtain the approximate frequency spectrum \cite{ref24}.
\begin{equation}
	{S(f)\approx}{\rm Rect}\left(\frac{f-\frac{f_{\rm L}+f_{\rm H}}{2}}{B}\right)\gamma e^{-j\frac{\pi T(f-f_{\rm L})^2}{B}},
	\label{9}
\end{equation}
where $\gamma=\sqrt{T/B}e^{j\pi/4}$. {{According to the definition that $\Gamma^{\rm R}(f)$ rotates signals by subtracting $\kappa(f)$ from the phase of $S(f)$ \cite{rref1}, $\kappa(f)$ is given by}}
\begin{equation}
	\kappa(f)=-\frac{T(f-f_{\rm L})}{B}+{\rm C},
	\label{10}
\end{equation}
where ${\rm C}$ is a real-valued constant and set to zero here. By substituting \eqref{8}, \eqref{9} and \eqref{10} into \eqref{7}, yielding
\begin{equation}
	\begin{aligned}
		g_{s}(t,f_{w})=\sigma\gamma e^{{ j}2\pi f_wt
		}\Gamma^{\rm S}(f_w) \text{sinc}\left[\pi\sigma t+ \frac{\sigma\angle \Gamma^{\rm S}(f_w)}{2 f_w}\right],
		\label{12}
	\end{aligned}
\end{equation}
where $\text{sinc}(x)=\sin(x)/x$. The kernel function \eqref{10} also applies to the PTFT of $y_m(t)$. According to \eqref{12}, $g_{y_m}(t,f_w)$ can be represented as
\begin{equation}
	\begin{aligned}
		g_{y_{m}}(t,f_{w})=&\sum_{k=1}^{K}\left[b_k\sigma\gamma e^{{ j}2\pi f_w\epsilon_m
		}\text{sinc}(\pi\sigma\epsilon_m)\right]+g_{n_m}(t,f_w),
		\label{13}
	\end{aligned}
\end{equation}
where $\epsilon_m=t-\tau^k+\angle\Gamma^{\rm S}(f_w)/(2\pi f_w)  -\tau_m(\theta_k)$. Then, a new array data $\widetilde{\pmb{{y}}}({t,f_{w}})$ can be formulated based on $g_{y_m}(t,f_w),\forall m$. 
\begin{equation}
	\begin{aligned}
		\widetilde{\pmb{{y}}}({t,f_{w}})=&\left[g_{y_1}(t,f_{w}),g_{y_2}(t,f_{w}),\cdots,g_{y_M}(t,f_{w})\right]^{\top}\\
		=&\sum_{k=1}^{K}{\text{diag}(\pmb{\rho}_{k})}\pmb{a}({f_{w}},\theta_k) \phi_k+\widetilde{\pmb{{n}}}(t,f_w),
		\label{17}
	\end{aligned}
\end{equation}
where ${{\pmb{\rho}_{k}=[\rho_{1,k},\cdots,\rho_{M,k}]^\mathsf{T}},\rho_{m,k}=|b_k|\sigma\sqrt{{T}/{B}}\text{sinc}(\pi\sigma\epsilon_m)}$, $\phi_k=\exp\left\{j\angle b_k\right. \left.+{{j}{\pi}/{4}+{ j}2\pi f_w\left[t-\tau^k+\angle\Gamma^{\rm S}(f_w)/(2\pi f_w)\right]}\right\}$, $\pmb{a}({f_w},\theta_k)$ is the steering vector associated with the frequency $f_w$, and $\widetilde{\pmb{{n}}}(t,f_w)=\left[g_{n_1}(t,f_w),g_{n_2}(t,f_w),\cdots,g_{n_M}(t,f_w)\right]^{\top}$. 
Based on \eqref{3} and \eqref{17}, the output SNRs of FFT and PTFT can be respectively expressed as
\begin{equation}
	\begin{aligned}
		\text{SNR}^{\rm FFT}=10\log\frac{|b_k|^2T}{|N_m(f)|^2B}.
		\label{14}
	\end{aligned}
\end{equation}
\begin{equation}
	\begin{aligned}
		\text{SNR}^{\rm PTFT}=10\log\frac{|b_k|^2\sigma^2\text{sinc}^2(\pi\sigma\epsilon_m)T}{|g_{n_m}(t,f_w)|^2B}.
		\label{15}
	\end{aligned}
\end{equation}

Usually, to attain reliable DOA estimation, the array data is constructed by samplings located at the {time-frequency} ridge {for their good concentration of signal energy \cite{ref17,ref20,ref21}. For the {time-frequency} ridge in \eqref{15}, we have $\epsilon_m=0$.} By setting $\Gamma^{\rm S}(f)=1$, the {time-frequency} ridge of the first sensor is moved to $t=\tau^k$. Let us collect the data at the {time-frequency} index $(\tau^k,f_w)$ for all sensors, then $\epsilon_m\neq0$ and $\text{sinc}(\pi\sigma\epsilon_m)=\text{sinc}[-\pi\sigma\tau_m(\theta_k)]\neq1$ except for $m=1$. {Upon exploiting Taylor series to approximate $\text{sinc}(\pi\sigma\epsilon_m)$,} we obtain $\text{sinc}(\pi\sigma\epsilon_m)=1-[\pi\sigma\tau_m(\theta_k)]^2/6+o\{\left[\pi\sigma\tau_m(\theta_k)\right]^4\}$. Thus, an appropriate $\sigma$ satisfying \eqref{24} can ensure that the amplitude error $\delta$ is within a reasonable range. 
\begin{equation}
	\begin{aligned}
	|\rho_{m,k}-\rho_{1,k}|\approx|b_k|\sigma\sqrt{{T}/{B}}[\pi\sigma\tau_m(\theta_k)]^2/6\leq \delta,\forall m.
	\label{24}
\end{aligned}
\end{equation}
{Moreover, it can be observed by \eqref{15} that $\text{SNR}^{\text{PTFT}}$ increases with $\sigma$, which theoretically suggests that PTFT can provide signal processing gain. Numerically,  
Fig. \ref{fig1} gives the mean} SNRs for both FFT and PTFT with different values of $\sigma$ and $M=16$ sensors. The input SNR is defined as $10\log\left\{|b_k|^2/\mathbb{E}[|n(t)|^2]\right\}$. It can be observed from Fig. \ref{fig1} that the output SNRs are fully overlapping when $\sigma=1{/T_\mathsf{all}}$ Hz, indicating that PTFT fails to yield a gain in this case. As $\sigma$ increases, the output SNR of PTFT exceeds that of FFT gradually, and the SNRs of all sensors are nearly consistent, which means $\delta$ is under an acceptable level. {Accordingly, the amplitudes $\rho_{m,k},\forall m$ in $\pmb{\rho}_k$ are approximately equal and are recorded as $\rho_k$, yielding}
\begin{equation}
    \begin{aligned}
        {\widetilde{\pmb{{y}}}({\tau^k,f_{w}})\approx\sum_{k=1}^{K}\pmb{a}({f_{w}},\theta_k) \rho_k\phi_k+\widetilde{\pmb{{n}}}(t,f_w).}
		\label{req6}
    \end{aligned}
\end{equation}
{Different $\tau^k$ implies that target signals are separable in the time-frequency domain, and thus samplings at different time-frequency ridges can be processed independently \cite{ref20}. We consider the case where $K$ signals overlap completely in the time-frequency domain, namely $\tau_1=\tau_2=\cdots=\tau^K=\tau$.} 
{Then, extracting the single-snapshot and multi-frequency samplings at $(\tau,f_w)$, the PTFT-based FD expression} $\widetilde{\pmb{{z}}}({\tau},f_w,f_{w^{\prime}})$ with $f_{w^{\prime}}=f_w+\Delta f$ is formulated as \eqref{req5}{, which includes $K$ real DOAs and $(K^2-K)$ artifact DOAs}.
\begin{figure*}[!t]
	\vspace*{4pt}
	\begin{equation}
		\begin{aligned}
			{\widetilde{\pmb{{z}}}(\tau,f_w,f_{w^{\prime}})}&{=\widetilde{\pmb{{y}}}({\tau,f_{w}+\Delta f})\odot \left[\widetilde{\pmb{{y}}}({\tau,f_{w}})\right]^\ast}
			\\&{={\underbrace{\sum_{k=1}^{K}\pmb{a}({\Delta f},\theta_{k})\rho_k^2\phi_{k,f_{w^\prime}}\phi_{k,f_{w}}^\ast}_{K \text{ self terms}}}	
			+{\underbrace{{\sum_{k_{1}=1}^{K}\sum_{\substack{k_2=1,\\k_{2}\neq k_1}}^{K}}\left[\pmb{a}({f_{w^\prime}},\theta_{k_{2}})\odot\pmb{a}^\ast({f_{w}},\theta_{k_{1}})\right]\rho_{k_2}\rho_{k_1}\phi_{k_2,f_{w^\prime}}\phi_{k_1,f_{w}}^\ast}_{(K^2-K) \text{ cross terms}}}
			+{\underbrace{\widetilde{\pmb{n}}(f_w,f_{w^{\prime}})}_{\text{noise terms}}}.}
			\label{req5}
		\end{aligned}
	\end{equation}
    
	\hrulefill
    
	\vspace{-0.6cm}
\end{figure*}
By replacing $\pmb{z}(f_1,f_2)$ in \eqref{req2} with $\widetilde{\pmb{{z}}}({\tau},f_w,f_{w^{\prime}})$, the framework of {the PTFT-based CFD} is established, which can be solved by {the CVX toolbox \cite{ref25}.}
{Each frequency pair $\{f_w,f_{w^{\prime}}\}$ with $f_w=f_{\rm L}+(w-1)f_{\rm step}$ can estimate a DOA spectrum for $w=1,2,\cdots,W$, where $W= \lfloor ({f_{\rm H}-f_{\rm L}}-\Delta f)/{f_{\rm step}}\rfloor$ and $f_{\rm step}$ is the frequency increment.}
\begin{figure}[H] 
	\centering
	\vspace{-0.6cm} 
	\subfloat[]{\includegraphics[width=0.5\linewidth]{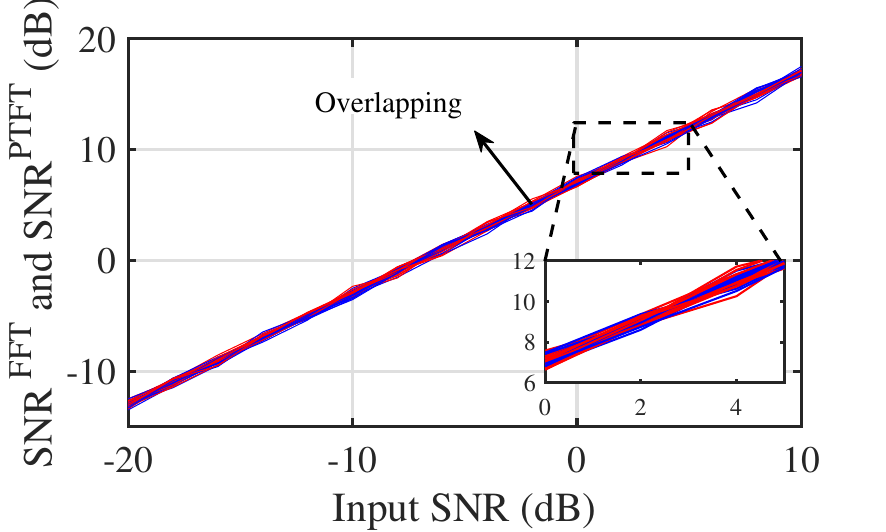}%
		\label{fig_1a}}
	\hspace{-4mm}
	\subfloat[]{\includegraphics[width=0.5\linewidth]{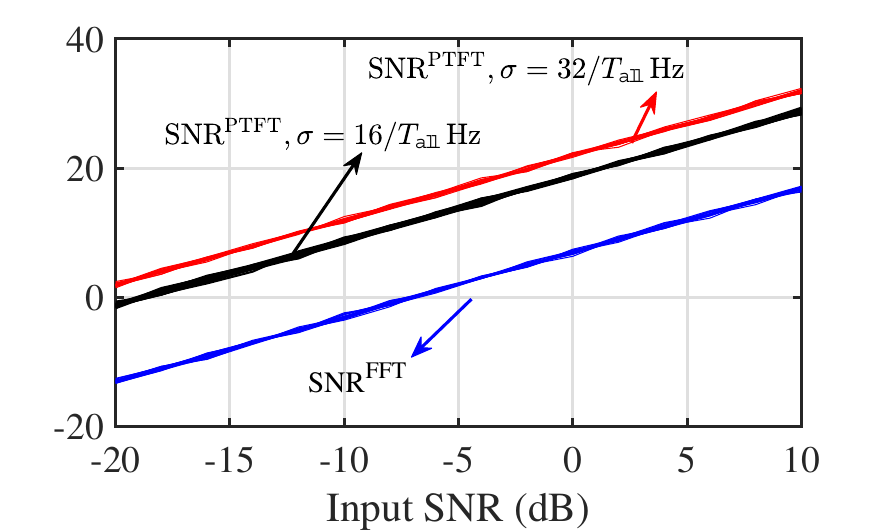}%
		\label{fig_1b}}
	\caption{$\text{SNR}^{\rm FFT}$ and $\text{SNR}^{\rm PTFT}$ vs. input SNR operating at (a) $\sigma=1{/T_\mathsf{all}}$ Hz, (b) $\sigma=16{/T_\mathsf{all}}$ Hz and $\sigma=32{/T_\mathsf{all}}$ Hz.}
	\label{fig1}
\end{figure}

\subsection{Coarse-to-Fine {Histogram Statistics Scheme}}
{Attempting to address the problem of pseudo-peaks present in conventional incoherent averaging over DOA spectra, we conceive a coarse-to-fine {histogram statistics} scheme, as shown in Fig. \ref{fig2}}. At first, we fetch $K^2$ DOAs from $w$th DOA spectrum searching and constitute $\hat{\Theta}(f_w)=\{\hat{\theta}_{f_w,k^\prime}|k^\prime=1,2,\cdots,K^2\}$, including {$K$} real DOAs and {$(K^2-K)$} artifact DOAs. By repeating $W$ times, a total collection $\hat{\Theta}=\hat{\Theta}(f_1)\cup\hat{\Theta}(f_2)\cup\cdots\cup\hat{\Theta}(f_W)$ with $(K^2\times W)$ elements is generated. Subsequently, {harnessing the feature that the real DOAs remain consistent while the artifact DOAs vary with frequency $f_w$ \cite{refr1,ref12,ref13},} a coarse {histogram statistics is executed to determine the initial $K$ bins}, which counts the elements in $\hat{\Theta}$ falling into bins of $\left[-90^\circ:\zeta:90^\circ\right]$ {and selects $K$ bins $\{\left[{B}_k,{B}_k+\zeta\right]|k=1,2,\cdots,K\}$ with the first $K$ maximum counts.} ${B}_k$ denotes the left endpoint of {the} $k$th bin {and $\zeta\in(0^\circ,180^\circ]$ is the bin width. A larger $\zeta$ allows for greater tolerance of DOA perturbations present in the DOA spectra at different $f_w$ but increases averaging error and reduces angular resolution. Conversely, a smaller $\zeta$ is insufficiently robust to perturbations. Thus, $\zeta=2^\circ$ is chosen as a trade-off.}


When the {real} DOAs are {near interval} endpoints, the intervals inferred from the coarse {histogram statistics} become biased. In this regard, a fine {histogram statistics is executed to enable fine tuning}. Based on the initial $k$th bin, it redraws three intervals {by extending bins at both ends}, i.e., $\mathbb{B}_{k,1}=\left[{ B}_k-{\zeta}/{2},{B}_k+{\zeta}/{2}\right]$, $\mathbb{B}_{k,2}=\left[{B}_k,{ B}_k+\zeta\right]$ and $\mathbb{B}_{k,3}=\left[{ B}_k+{\zeta}/{2},{ B}_k+{3\zeta}/{2}\right]$. 
Here, we define an indicator function as
\begin{equation}
		{\mathds{1}_{k,i}(\hat{\theta}_{f_w,k^\prime})=}
	\left\{
	\begin{aligned}
		1,\ &\hat{\theta}_{f_w,k^\prime}\in \mathbb{B}_{k,i},\\
		0,\ &{\rm otherwise}.\\
	\end{aligned}
	\right.
	\quad{i=1,2,3.}
    \label{22}
\end{equation}
Next, the count in the interval $\mathbb{B}_{k,i}$ is calculated as
\begin{equation}
    \chi_{k,i}=\sum_{w=1}^{W}\sum_{k^\prime=1}^{K^2}\mathds{1}_{k,i}(\hat{\theta}_{f_w,k^\prime}).
	\label{23}
\end{equation}
The interval $\mathbb{B}_{k,i^\prime}$ with $i^\prime={\arg\max}_{i}\,\{\chi_{k,i}|i=1,2,3\}$ is determined as the final $k$th bin. Finally, the  $k$th estimated DOA $\hat{\theta}_k$ is computed by 
averaging elements in $\hat{\Theta}\cap\mathbb{B}_{k,i^\prime}$. {In contrast
to incoherent averaging over DOA spectra, averaging over DOA estimates avoids pseudo peaks
due to DOA perturbations and can leverage the random distribution of DOA estimation errors across
frequency $f_w$ to improve estimation accuracy. In supplementary materials, we provide the overall flowchart of the proposed PTFT-based HS-CFD algorithm.}
\begin{figure}[H]  
	\centering  
	\vspace{-0.35cm} 
	\includegraphics[width=0.7\linewidth]{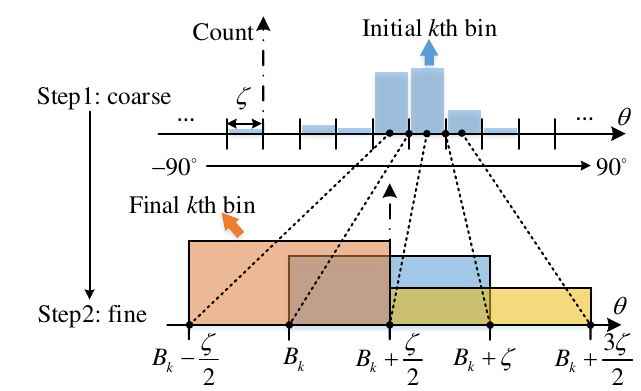}
	\caption{Diagram of the coarse-to-fine {histogram statistics} scheme.}
	\label{fig2}
	\vspace{-0.2cm}
\end{figure}
 
\section{Simulation Results}
\label{simulation}
We consider a {sparse uniform linear array} with $M=16$ sensors and spacing $d=3.75$ m in an underwater scenario, where  $c=1500$ m/s. The strengths of two target signals from $\theta_1$ and $\theta_2$ are assumed equal\cite{ref13}. The frequency bandwidth $B=10$ kHz with $f_{\rm L}=10$ kHz and $f_{\rm H}=20$ kHz. Both the pulse duration and the overall signal length are $T={T_\mathsf{all}}=1$ s. The remaining parameters are set as {$\sigma=32{/T_\mathsf{all}}$ Hz, $\zeta=2^\circ$ and $\mu=0.1$.} The DOA searching range is divided into $1801$ grids, i.e., $[-90^\circ:0.1^\circ:90^\circ]$. The root mean-squared-error (RMSE) is defined as

\begin{equation}
	\text{RMSE} = \sqrt{\frac{1}{{I}K}\sum_{i=1}^{{I}}\sum_{k=1}^{K}(\theta_{i,k}-\hat{\theta}_{i,k})^2},
	\label{21}
\end{equation}
where $\theta_{i,k}$ and $\hat{\theta}_{i,k}$ are the true and estimated values, respectively, while {$I$} denotes the total number of Monte-Carlo experiments. 

Fig. \ref{fig3} depicts the DOA spectra calculated by {the FFT-based CFD \cite{ref13} and PTFT-based CFD}. It can be observed that {the PTFT-based CFD} can distinguish $\theta_1$ and $\theta_2$ at low SNRs compared to CFD. Nevertheless, slight fluctuation around $\theta_1$ and $\theta_2$ {{present} in Fig. \ref{fig3} \subref{fig_3b} can influence the subsequent incoherent {averaging, as indicated in the zoomed-in figure of Fig. \ref{fig4} \subref{fig_4a}}.} The minor perturbation leads to {pseudo peaks}, while the result of {the first histogram statistics} in Fig. \ref{fig4} \subref{fig_4b} is less perturbed {due to the coarse grid processor}. Following this, the {second histogram statistics fine tunes the intervals and} yields $\hat{\theta}_1=-0.0647^\circ$ and $\hat{\theta}_2=15.1092^\circ$, {which validates that the coarse-to-fine {histogram statistics} scheme is capable of mitigating the perturbation without the loss of DOA estimation accuracy.}
\begin{figure}
	\centering
	\vspace{-0.35cm} 
	\subfloat[]{\includegraphics[width=0.5\linewidth]{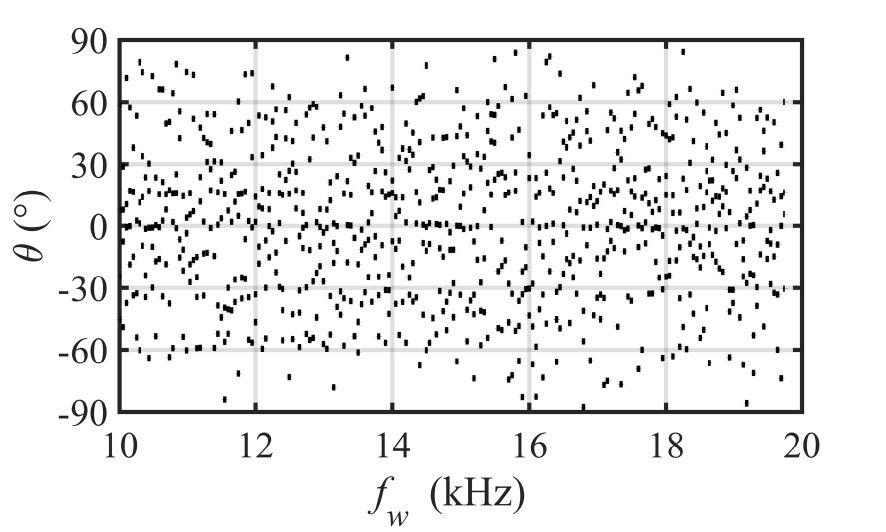}%
		\label{fig_3a}}
	\subfloat[]{\includegraphics[width=0.5\linewidth]{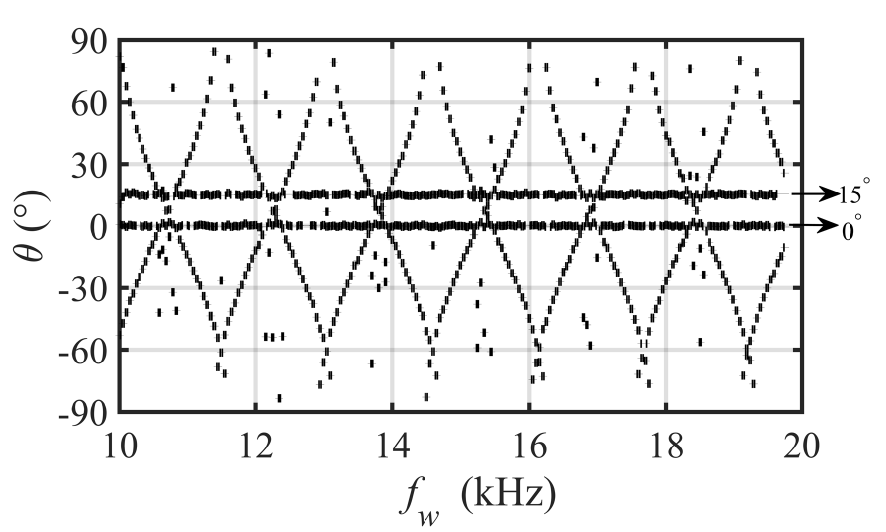}%
		\label{fig_3b}}
	\caption{DOA spectrum vs. $f_w$ under $\text{SNR}=-14$ dB with $\theta_1=0^\circ$ and $\theta_2=15^\circ$ {by (a) FFT-based CFD and (b) PTFT-based CFD}.}
	\label{fig3}
\end{figure}
\begin{figure}
	\centering
	\vspace{-0.6cm} 
	\subfloat[]{\includegraphics[width=0.5\linewidth]{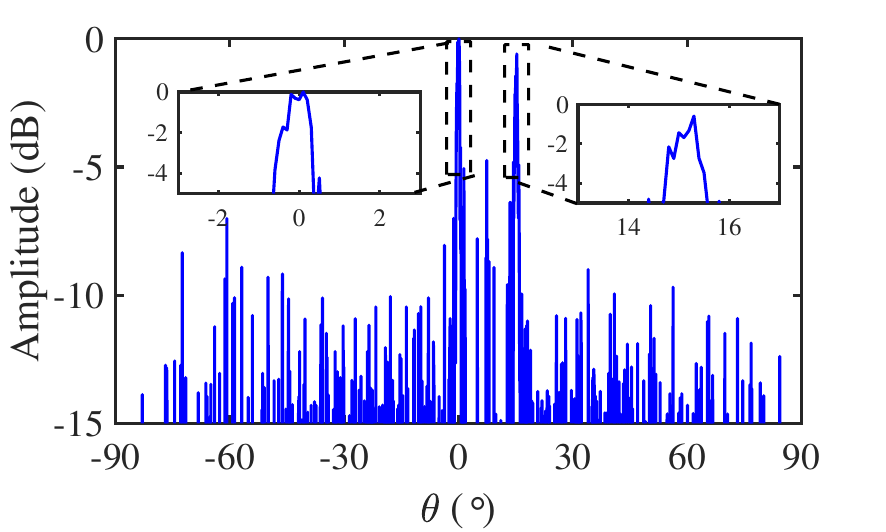}%
		\label{fig_4a}}
	\subfloat[]{\includegraphics[width=0.5\linewidth]{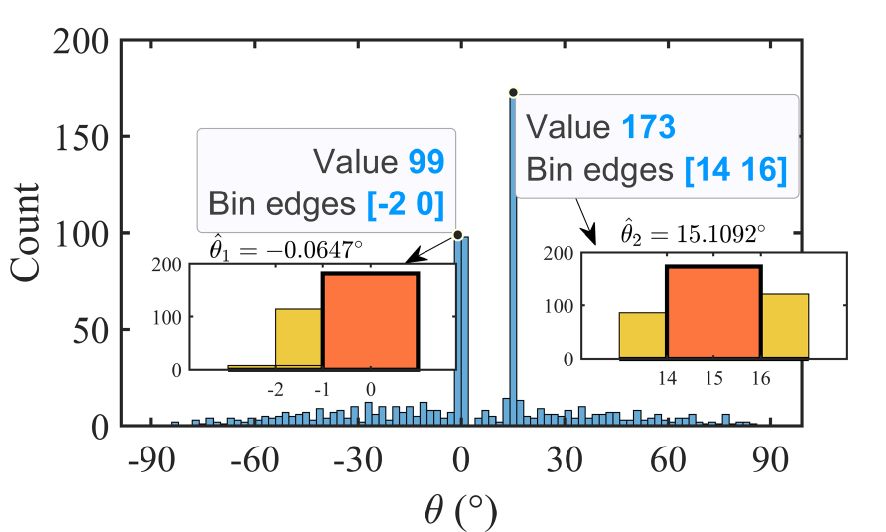}%
		\label{fig_4b}}
	\caption{Processing results of (a) incoherent {averaging} and (b) {coarse-to-fine histogram statistics scheme}.}
	\label{fig4}
	\vspace{-0.4cm}
\end{figure}

Fig. \ref{fig5} compares the robustness, accuracy and resolution of {the FFT-based algorithms, including FD-CBF \cite{ref11}, FD-MUSIC \cite{ref12}, CFD \cite{ref13}, the multi-snapshot STFT-based algorithms, including FD-ANM and FD-ANM-spectral-averaging (FD-ANM-SA) \cite{refr1}, and the proposed PTFT-based HS-CFD algorithm}, which considers $\theta_1=0^\circ$ with different $\theta_2$ settings under the input $\text{SNR}=-14$ dB. Fig. \ref{fig5} plots $\hat{\theta}_1$ and $\hat{\theta}_2$ of four algorithms {, and the results of FD-ANM and FD-ANM-SA are detailed in supplementary materials}. The black solid lines denote the true values of $\theta_1$ and $\theta_2$. {It can be observed that {the proposed algorithm surpasses FD-MUSIC and CFD in robustness due to the higher output SNR of PTFT}.} Based on the zoomed-in views in Figs. \ref{fig5} \subref{fig_5a} and \ref{fig5} \subref{fig_5d}, we find that {the proposed algorithm provides more accurate DOA estimates than FD-CBF.} Moreover, as indicated by the arrows in Figs. \ref{fig5} \subref{fig_5a} and \ref{fig5} \subref{fig_5d}, {the proposed algorithm exhibits a minimum resolvable angle of approximately $4^\circ$, lower than FD-CBF}. {This is because {compressive beamforming} can attain high-resolution \cite{ref13} and {the proposed algorithm} inherits it.} {These observations verify that {the proposed} algorithm is capable of achieving better robustness, accuracy and resolution performance compared to other solutions.}

{Fig. \ref{fig6} investigates the RMSE performance of {the FFT-based FD algorithms, the STFT-based FD algorithms, and the PTFT-based FD algorithms} under different input SNRs.} It can be concluded that {the PTFT-based FD-CBF, FD-MUSIC, and HS-CFD} are capable of attaining SNR gains and {enhancing DOA estimation} accuracy compared to {other} paradigms. Specifically, the {estimation errors of the FFT-based FD-CBF} and FD-MUSIC {converge to} $0.7^\circ$ when input SNRs are about $-12$ dB and $-8$ dB, while the {PTFT-based FD-CBF and FD-MUSIC} can attain this at {about $-28$ dB and $-24$ dB. {Their converged values of DOA estimation errors are affected by a damped oscillation for coherent arrivals} \cite{ref13}. The estimation accuracy of both FD-ANM-SA and CFD degrades due to pseudo peaks from incoherent averaging. FD-ANM exhibits lower estimation errors around $0.5^\circ$, while the performance of FD-ANM is limited by the STFT's inability to effectively aggregate energy from non-stationary signals.} {Compared to these counterparts, the {proposed algorithm} can offer the best RMSE performance, which remains around $0.04^\circ$ from $-16$ dB.} 
{The superior performance of the proposed algorithm is primarily due to three factors: its ability to effectively handle single-snapshot and coherent signal scenarios using compressed sensing, the improvement in signal processing gain achieved by PTFT, and the ability of the coarse-to-fine scheme to effectively reduce DOA perturbation and leverage its randomness to decrease estimation error.}

\begin{figure}
	\centering
	\vspace{-0.5cm} 
	\subfloat[]{\includegraphics[width=0.5\linewidth]{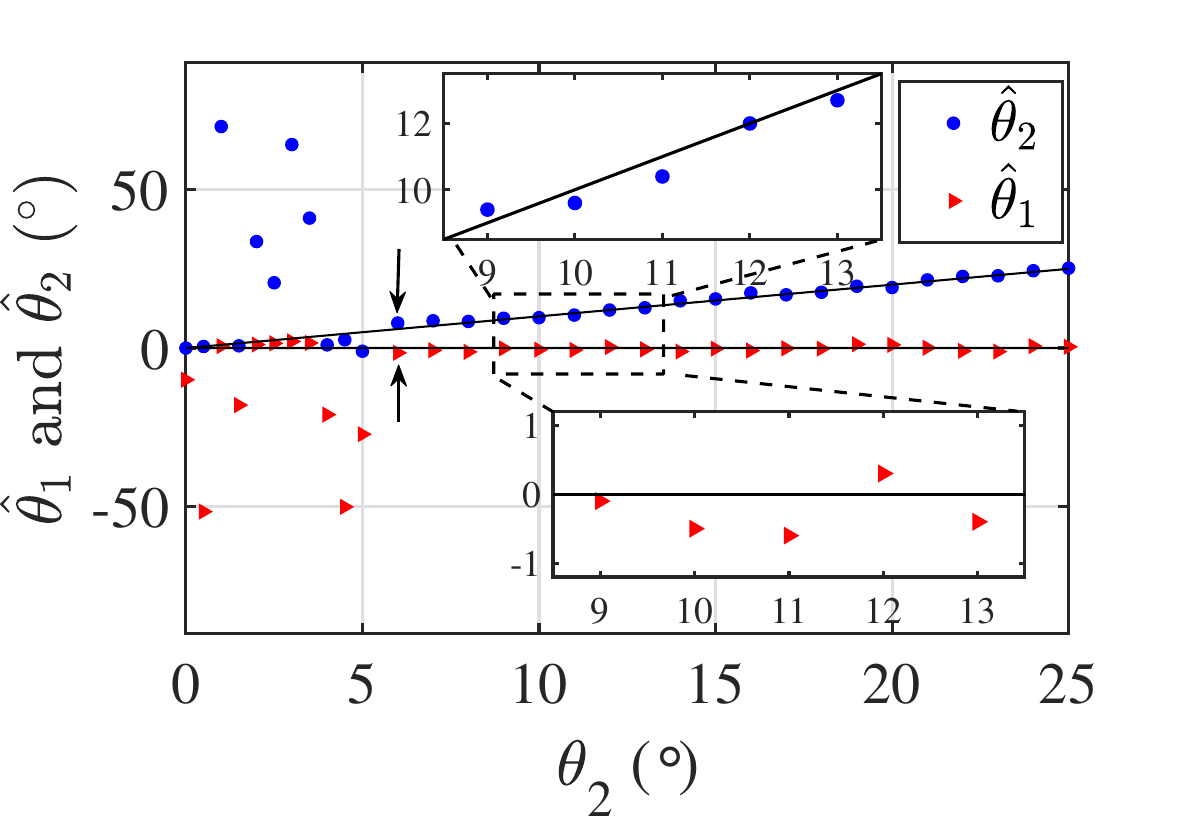}%
		\label{fig_5a}}
	\subfloat[]{\includegraphics[width=0.5\linewidth]{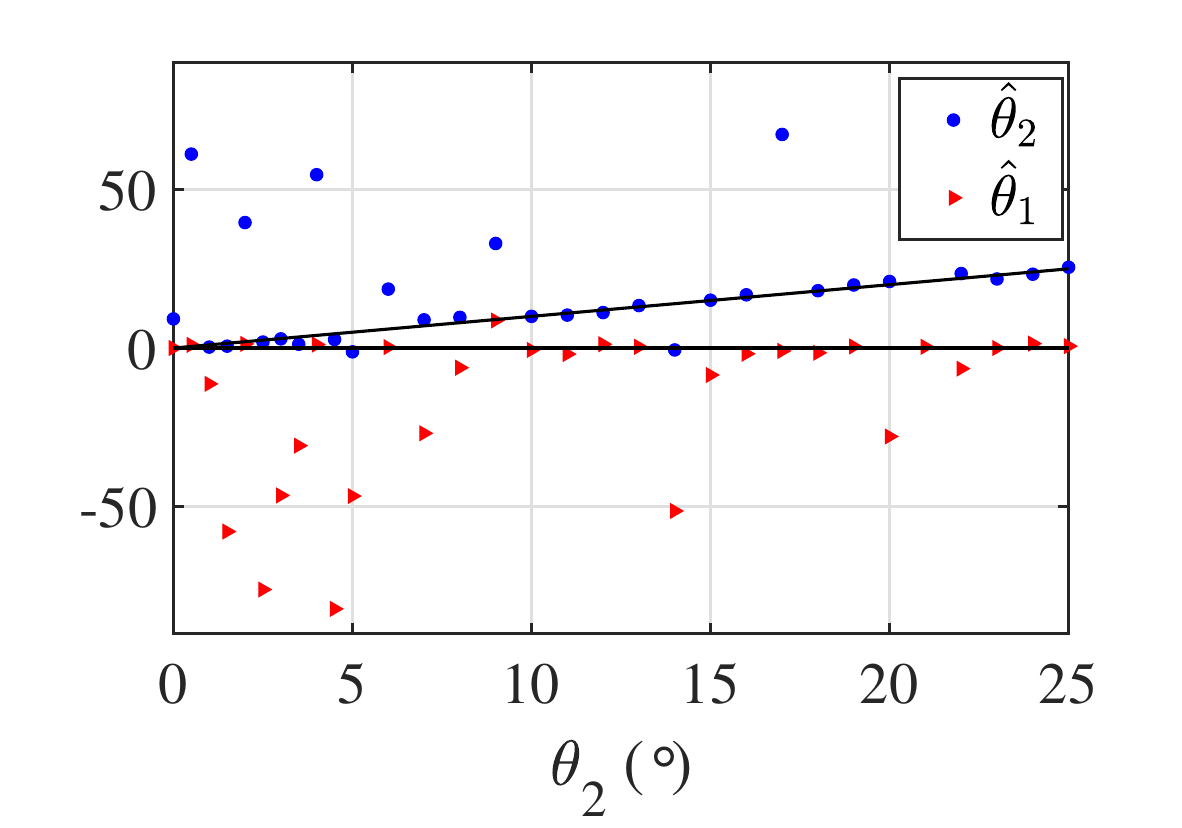}%
		\label{fig_5b}}
	\vspace{-0.4cm} 
	\subfloat[]{\includegraphics[width=0.5\linewidth]{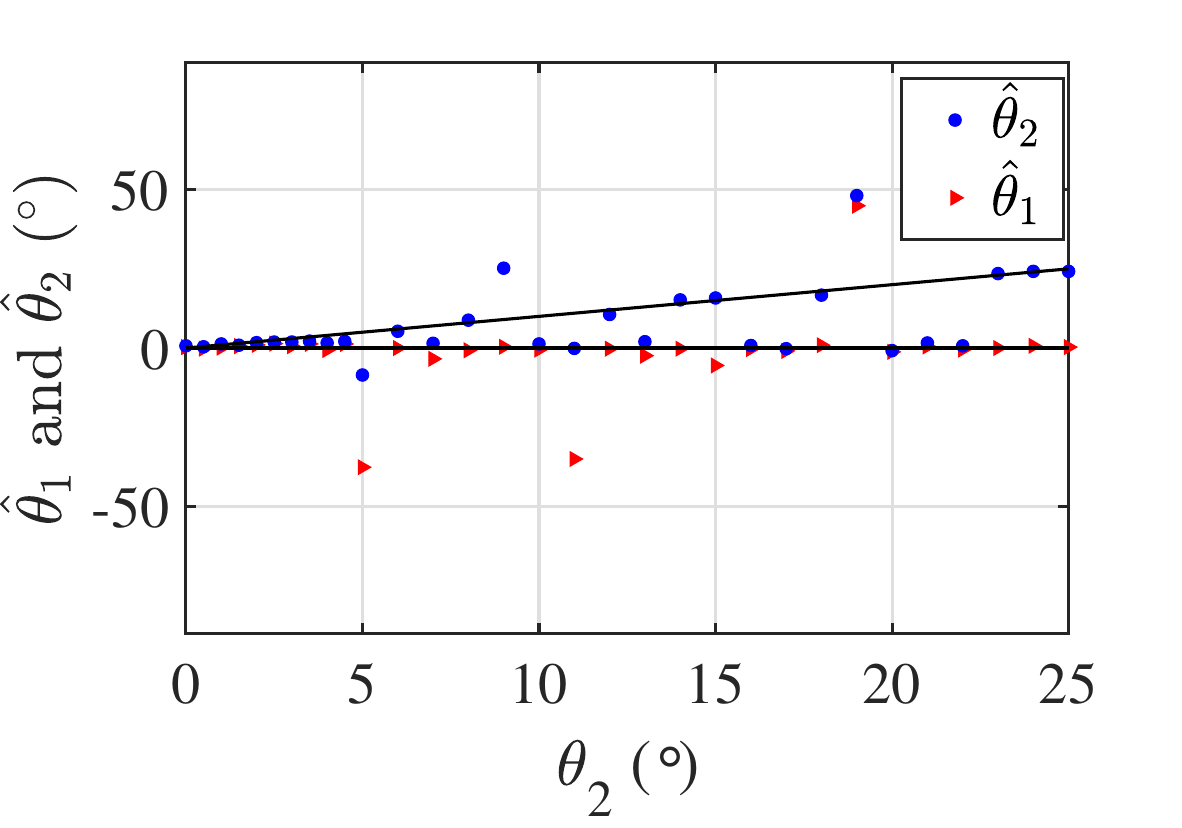}%
		\label{fig_5c}}
	\subfloat[]{\includegraphics[width=0.5\linewidth]{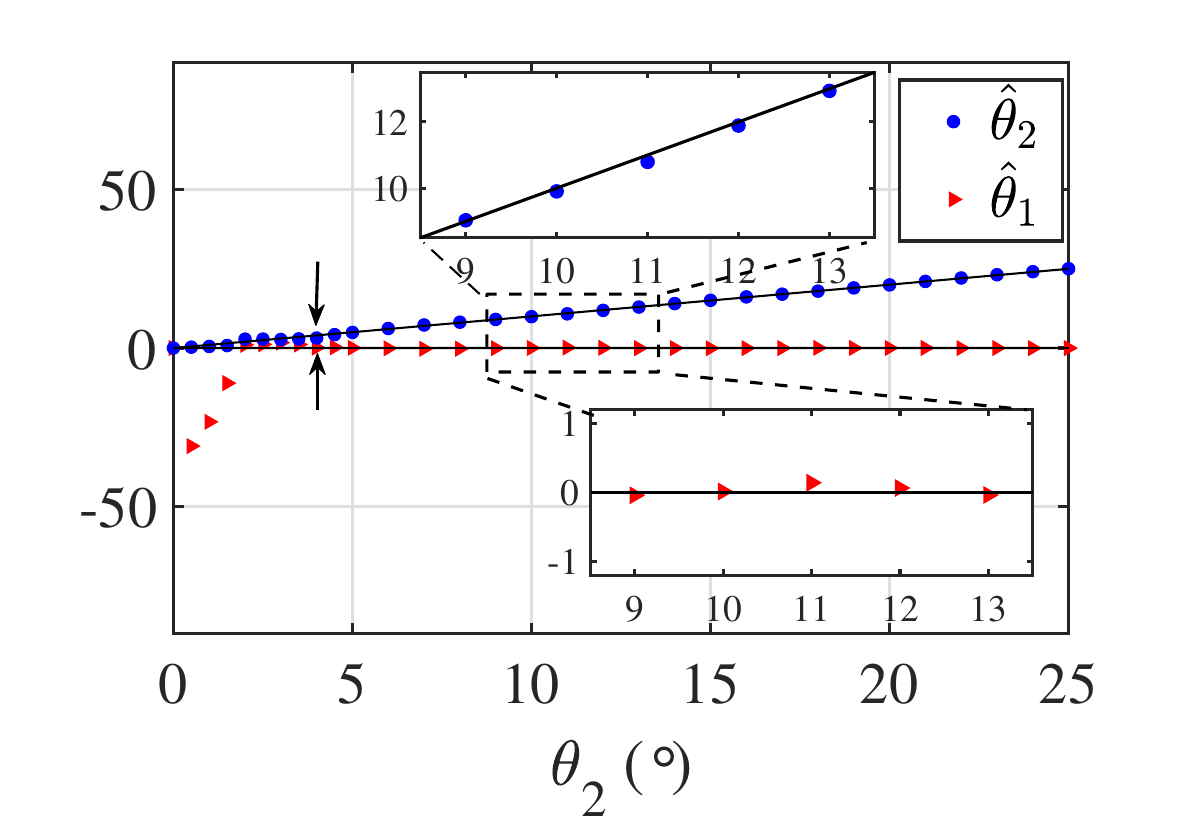}%
		\label{fig_5d}}
	\caption{Estimated DOAs $\hat{\theta}_1$ and $\hat{\theta}_2$ of { (a) FD-CBF}, (b) FD-MUSIC, (c) CFD and (d) {the proposed algorithm} with $\theta_1=0^\circ$ and $\theta_2=[0^\circ:0.5^\circ:5^\circ,6^\circ:1^\circ:25^\circ]$.}
	\vspace{-0.4cm}
	\label{fig5}
\end{figure}

\begin{figure}
	\centering  
	\includegraphics[width=0.8\linewidth]{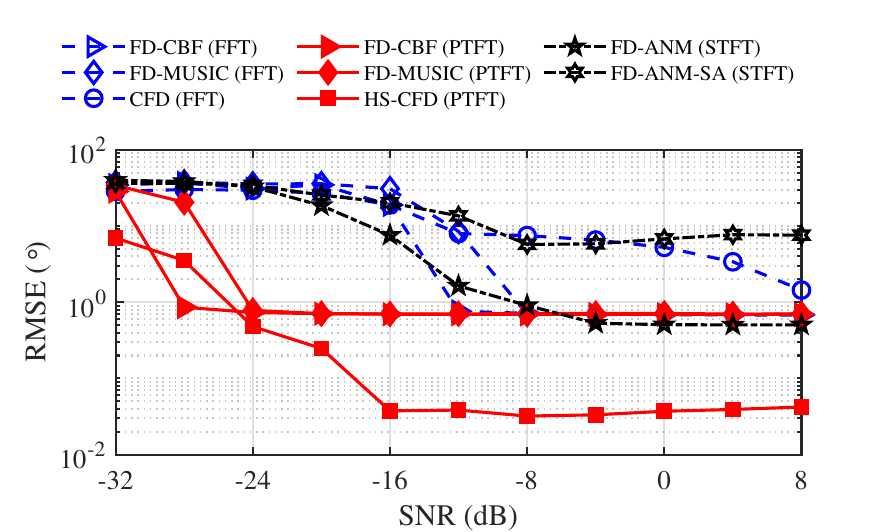}
	\caption{RMSE vs. input {$\text{SNR}=\left[{-32}:4:8\right]$ dB} with $\theta_1=0.78^\circ$ and $\theta_2=15.23^\circ$.}
	\label{fig6}
    \vspace{-0.4cm}
\end{figure}

\section{Conclusions}
\label{conclusion}
An unambiguous DOA estimation algorithm has been proposed for {sparse uniform linear arrays} with non-stationary broadband signals. Explicitly, a PTFT-based {FD} scheme has been proposed, in which {the signal energy is better aggregated compared to the conventional FD.}
Then, the {compressive beamforming} has been {combined} to strike high-resolution DOA estimation. Although the {PTFT-based CFD} has benefited from {signal processing gain of PTFT}, the perturbation in {CFD} can lead to inaccurate DOA estimation due to incoherent {averaging} operations. Hence, we {have further conceived} a coarse-to-fine {histogram statistics} scheme to alleviate perturbation while {maintaining the DOA estimation accuracy.} Simulation results have shown that the proposed ambiguity-free algorithm can attain robust and precise DOA estimation with high resolution.}


\balance

\end{document}